# Моделирование генных регуляторных сетей у дрозофилы с учетом модульной структуры гена

**Е.М. Мясникова[1], А.В. Спиров[2]**

[1]*Санкт-Петербургский Политехнический университет Петра Великого;*

[2]*Институт Эволюционной Физиологии и Биохимии им. Сеченова, СПб; Университет Стони Брук, США.*

*E-mail: myasnikova@spbcas.ru*

Цель работы состоит в развитии подхода к моделированию генных регуляторных сетей с учетом модульной структуры гена. Модель строится на основе экспериментальных данных и последующей экспериментальной верификации. Особенность подхода состоит в том, в модель включают дополнительные молекулярно-биологические детали, что более реалистично отражает механизмы контроля генной активности. Модельным объектом служат гены сегментации *hb* (*hunchback*) и *Kr* (Kruppel) семейства gap у дрозофилы.

Особое внимание уделено разработке методов оценки чувствительности модели к изменениям начальных условий и входных факторов. Основные цели методов – выделение факторов в наибольшей степени, регулирующих формирование характерных признаков паттернов, имеющих важное биологическое значение.

## 1. Введение

В последние годы достигнут значительный прогресс в использовании математического моделирования для выявления механизмов сложных регуляторных взаимодействий, определяющих экспрессию генов. Мы используем подход к моделированию с подгонкой к экспериментальным данным [1], при котором на вход модели экспрессии гена-мишени (транскрипции для РНК и трансляции для белка) подаются пространственные паттерны регуляторов его транскрипции, а получающиеся на выходе модели пространственные паттерны подгоняются к экспериментальным данным по экспрессии гена-мишени (обычно дикого типа, ДТ, т.е. в немутантных организмах). Тем самым, определяются оптимальные значения параметров модели и подбираются наиболее подходящие элементы модели и их взаимодействия (см. обзор [2]). Затем модель может уточняться для того, чтобы воспроизводить паттерны экспрессии при мутациях или других экспериментальных вмешательствах.

В раннем развитии мушки дрозофилы процесс формирования паттернов вдоль главной оси тела эмбриона очень хорошо изучен [3-4]. Небольшая генетическая сеть взаимных регуляторов, транскрипционных факторов (ТФ), (~15 генов сегментации) закладывает на основе исходных материнских факторов (пространственных градиентов концентраций) основу для будущего строения тела насекомого [5-6]. Ввиду этого ансамбль генов этого семейства служит популярным объектом математического моделирования и большая часть усилий по созданию моделей с целью лучшего понимания динамики генной сети, предпринималась именно в отношении этой системы. Главным образом, моделирование осуществлялось с помощью уравнений реакции-диффузии, причем взаимная регуляция генов была представлена линейно в виде матрицы взаимодействий (gene circuits или коннекционистские модели) [7-11]. Подобное представление является существенно упрощенным, поскольку не допускает синергетических или ко-факторных (нелинейных) эффектов между несколькими ТФ, при этом каждый ген рассматривается как отдельная мишень для отдельного ТФ.

Впервые наш новый подход к моделированию экспрессии гена независимо с разных ЦРМ, допускающий нелинейные взаимодействия генов, был успешно применен для описания формирования паттерна экспрессии гена hunchback (hb) в эмбрионе дрозофилы (цикл деления 14А) [12]. Этот ген является ключевым в ранней сегментации дрозофилы [13-14] и других насекомых [15-16]. Экспрессия *hb* контролируется тремя различными ЦРМ, расположенными перед кодирующей областью [17-19]. К тому же, как и у многих других генов, транскрипция hb осуществляется с двух промоторов P1 и P2. Они производят несколько отличающиеся РНК-транскрипты, транслирующиеся в один и тот же белок Hb. С использованием метода компьютерного эволюционных вычислений [20] нами были найдены группы ТФ, ответственных за формирование паттернов обоих транскриптов, и силу регуляторных воздействий этих ТФ на экспрессию *hb*. Ген *hb* принадлежит семейству gap, и также как другие гены этого семейства, сначала экспрессируется в виде одного широкого домена (области экспрессии), который затем в процессе развития разделяется на несколько полос (пиков). Такая динамика формирования паттерна является следствием модулярности: разные ЦРМ вовлечены в этот процесс на разных стадиях развития.[21]

С целью воссоздания более реалистичной динамической картины формирования паттернов экспрессии в результате взаимодействия генов, мы идем дальше и включаем в модель второй ген того же семейства gap, *Krüppel* (*Kr*), следующий в каскаде генов сегментации вслед за *hb*. Экспрессия *Kr* контролируется двумя цис-регуляторными элементами: CD1 и CD2 [22-23]. Однако мы не обладаем информацией о дифференциальной экспрессии этого гена с двух разных ЦРМ. Экспрессия *Kr* локализуется в точно ограниченной области в центре эмбриона, границы которой контролируются в передней части морфогенетической активностью генов *bicoid* (*bcd*) и *hb* [15,23]. Среди других ТФ, регулирующих Kr, выделяют Knirps (Kni), который действует как репрессор (точнее, как репрессор активации Kr морфогеном Bcd) (см. [24] и ссылки), а также репрессор Giant (Gt).

В нашей модели динамики одного гена *hb* [12] показано, что для корректного воспроизведения и нормы, и мутантных по *Kr* паттернов требуется вовлечение минимум еще одного гена нуббина в минимальный ансамбль генов сети. Этот ген относят к, так называемым, gap-подобным (gap-like) генам и область его экспрессии располагается в задней части эмбриона. В настоящей работе мы ставим цель исследовать роль нуббина в контроле активности *hb* и *Kr* на основе двугенной модели.

Существенным упрощением действительности в нашей модели является то, что она описывает динамику только двух генов семейства gap, при этом остальные ТФ, включенные в модель предполагаются постоянными во времени внешними факторами. В дальнейшем развитии модели мы будем придавать ей более реалистичный характер, учитывая динамический характер формирования паттернов экспрессии всех генов ансамбля.

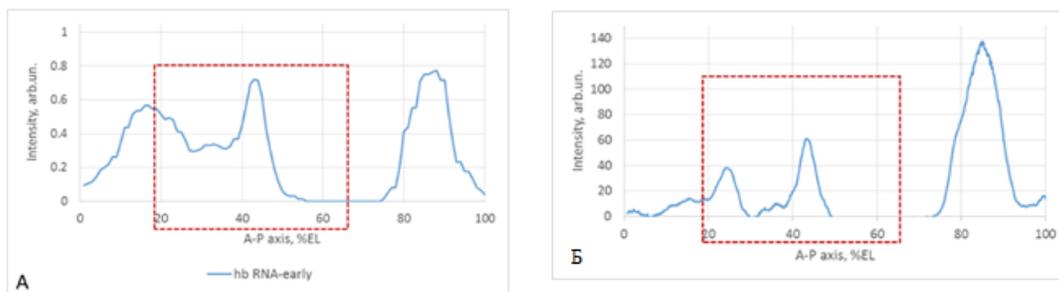

**Рисунок 1** — Экспериментальные паттерны экспрессии мРНК *hb*. (А) В начале цикла NC14 (используемые как начальные данные для (1).). (А) В середине цикла NC14.

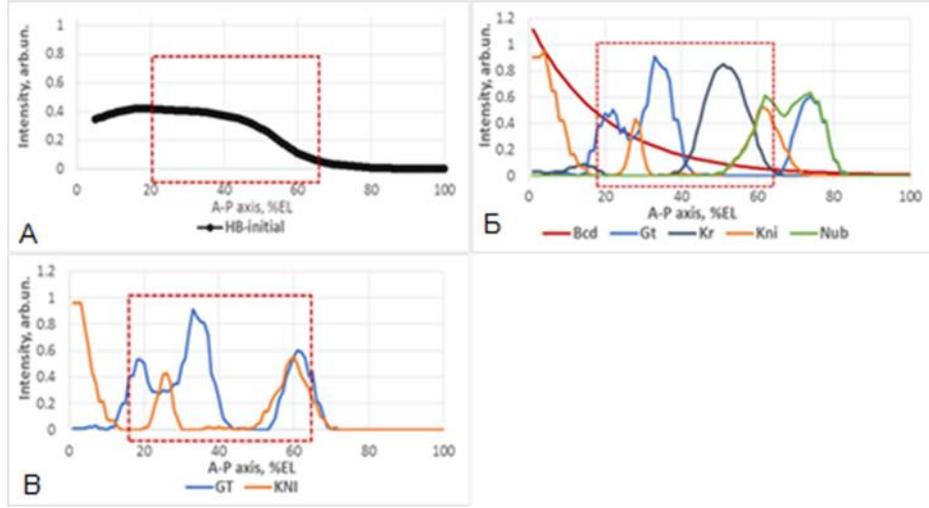

**Рисунок 2** — Экспериментальные паттерны экспрессии ТФ, используемые как начальные данные для (2). (А) Начальный паттерн *hb* в конце цикла NC13 из БД FlyEx. (Б) Паттерны ТФ семейства gap. (В) Экспрессия Gt и Kni в *Kr*- мутантах. Красная рамка очерчивает позиции на оси АП, в которых проводится моделирование

## 2. Модель

В работе [12] уравнения модели были представлены в самом общем виде, которые затем уточнялись методом ГД. Паттерн экспрессии мРНК представляется как распределение концентрации транскрипта данного гена по ядрам эмбриона (рисунок 1), в нашем случае в одномерной цепочке из 100 ядер вдоль его продольной оси.

Для обоих генов паттерн экспрессии промотора $K$ моделируется во времени и пространстве (ядрах эмбриона) уравнением относительно концентрации мРНК в -м ядре:

$$\partial P_i^{(K)}(t)/\partial t = R \cdot g[\Phi_K(TF_1, TF_2, \ldots, TF_K) + H] - \lambda P_i^{(K)}(t), \qquad (1)$$

в котором синтез РНК описывается первым членом — членом реакции. Здесь $g(x) = 2/(1 + e^{-x}) - 1$ при $x \geq 0$, $g(x) = 0$ при $x < 0$ – логистическая функция сигмоидного типа, $R$ – коэффициент синтеза, $H$ – порог синтеза, $\lambda$ — коэффициент распада. Вид функционала Ф задается в рамках модели, и представляет собой сумму в общем случае нелинейных членов от концентрации транскрипционных факторов (ТФ) – продуктов генов-регуляторов, $TF_k$. Среди них могут быть и моделируемые продукты генов (*hb* и *Kr*), и постоянные во времени, но распределенные в пространстве входные факторы, полученные экспериментально. Если модель включает экспрессию с нескольких ЦРМ (промоторов), то вид функционала Φ определяется для каждого ЦРМ отдельно. Концентрация всех ТФ распределена в пространстве, ее паттерны строятся по экспериментальным данным, коэффициенты силы воздействия каждого ТФ на моделируемый ген находятся оптимизацией с помощью генетического алгоритма (ГА). Выбор релевантных факторов осуществляется из следующего набора: материнского градиента Bcd и продуктов генов gap - Gt, Hb, Kni, Kr. Все эти факторы, как известно, задействованы в регуляции моделируемых генов *hb* и *Kr* [23]. Более детально принципы построения функционала Φ будут описаны ниже.

На следующем этапе РНК-транскрипты транслируются в белок, уравнение синтеза которого в нашей модели имеет вид реакции-диффузии:

$$\partial S_i(t)/\partial t = R_{pr}\left[P_i^{(1)}(t) + P_i^{(2)}(t)\right] - \lambda_{pr} S_i(t) + D_{pr} \cdot diff[S_i(t)] \qquad (2),$$

где член синтеза включает сумму обоих транскриптов. Границы областей экспрессии сглаживаются за счет члена диффузии. Экспериментальные паттерны продуктов генов семейства gap представлены на рисунке 2.

В модели *hb* акцент делается на экспрессии пика "PS4" в центральной части эмбриона, которая в эмбрионах дикого типа складывается из экспрессии промоторов P1 и P2. PS4 формируется к середине цикла деления 14A и имеет важное значение для нормального морфологического развития мухи. Известно что в эмбрионах нуль мутантных по *Kr* экспрессия PS4 значительно снижается, причем экспрессия с промотора P1 отсутствует, т.е. в мутантах эта область экспрессии полностью контролируется P2.

Мы ставим своей целью создать модель, способную правильно воспроизводить паттерны экспрессии *hb* и *Kr* на уровне мРНК и белка в ДТ и $Kr^-$ и $kni^-$ мутантах. В $Kr^-$ мутанте это достигается тем, что модель P1 при обнулении входных данных по экспрессии Kr приводит к исчезновению пика PS4.

### 3. Результаты моделирования

Подгонка модели к экспериментальным данным с целью получения оптимальных оценок параметров и структуры модели производится с помощью генетического алгоритма (ГА) [20]. Для конкретизации модели требуется найти оптимальный вид функционала Ф и оценки параметров. В результате применения ГА нами был определен следующий вид функционала для моделирования продукции *hb* с промоторов P1 и P2:

$$\Phi_{Hb}^{(1)}: \quad \alpha_{Hb}^{(1)} S_{Hb}(t) + \alpha_{Kr1}^{(1)} S_{Kr}(t) - \alpha_{Kr2}^{(1)} S_{Kr}^2(t) - \alpha_{Gt}^{(1)}[\text{Gt}] - \alpha_{Kni}^{(1)}[\text{Kni}] - \alpha_{Nub}^{(1)}[\text{Nub}]; \quad (3)$$

$$\Phi_{Hb}^{(2)}: \quad \alpha_{Hb}^{(2)} S_{Hb}(t) + \alpha_{Bcd}^{(2)} F([\text{Bcd}]; K_{Bcd}) - \alpha_{Kr}^{(2)} F(S_{Kr}(t); K_{Kr}) - \alpha_{Kni}^{(2)}[\text{Kni}] - \alpha_{Gt}^{(2)}[\text{Gt}] - \alpha_{Nub}^{(2)}[\text{Nub}]; \quad (4)$$

где $F([\text{TF}]; K_{TF}) = [\text{TF}]/(K_{TF} - [\text{TF}])$,

На входе ГА в качестве исходного вида $\Phi_{Hb}^{(1)}$ и $\Phi_{Hb}^{(2)}$ используются уравнения, полученные в статье [12], которые в результате несколько модифицируются к виду (3-4) за счет включения члена самоактивации *hb* (первый член). Необходимость самоактивации для правильного формирования границы домена описана, например, в [11,25].

Сценарий формирования паттерна P1 в (3) отражает двойственный характер регуляции hb фактором Kr: активация (положительное воздействие) при низких и ингибирование (отрицательное воздействие) при высоких концентрациях Kr [26]. Также уравнения (3-4) включают репрессию Kni с тем, чтобы ограничить область экспрессии hb справа от пика Kr (см. рисунок 3.). Еще один член репрессии в области активности *kni* необходим для сохранения границы области hb в *kni*-мутантах — эту роль в модели играет ТФ Nub [27-29].

Таким образом, экспрессия *hb* регулируется самим Hb, Kr и еще четырьмя ТФ – Gt, Kni, Bcd и Nub, рассматриваемыми как постоянные во времени входные факторы.

В настоящей работе модель дополнена уравнением, описывающим динамику гена *Kr*, который в модели [12] также полагался неизменяющимся входным фактором. Формирование паттерна мРНК *Kr* моделируется одним уравнением вида (1), где:

$$\Phi_{Kr}: \quad \beta_{Kr} S_{Kr}(t) + \beta_{Hb} S_{Hb}(t) + \beta_{Hb2} S_{Hb}^2 + \beta_{Hb,Bcd} S_{Hb}(t) \cdot [\text{Bcd}] - \beta_{Gt}[\text{Gt}] - \beta_{Kni}[\text{Kni}] - \beta_{Nub}[\text{Nub}], \quad (5)$$

включающим саморегуляцию Kr, нелинейную регуляцию Hb, и линейную факторами Bcd, Gt, Kni и Nub.

Тем самым, в уравнении (2) вместо суммарной концентрации $P_i^{(1)}(t) + P_i^{(2)}(t)$ стоит общая концентрация транскрипта *Kr*, вычисляемая по формулам (1) и (5).

Динамика формирования пространственного паттерна концентрации продуктов генов (белка) $S_{Hb}(t)$ и $S_{Kr}(t)$ описывается уравнениями вида (2). Уравнения определяют экспрессию в каждом ядре.

Результаты моделирования представлены на рисунке 3.

Математически продукция мРНК генов определяется первым членом уравнения вида (1), где функционал Ф используется как аргумент сигмоиды $g(\cdot)$, обладающей свойством насыщения при высоких и отрицательных значениях аргумента. Тем самым продукция генов осуществляется только в пространственном и временном интервале, где Ф принимает положительные значения.

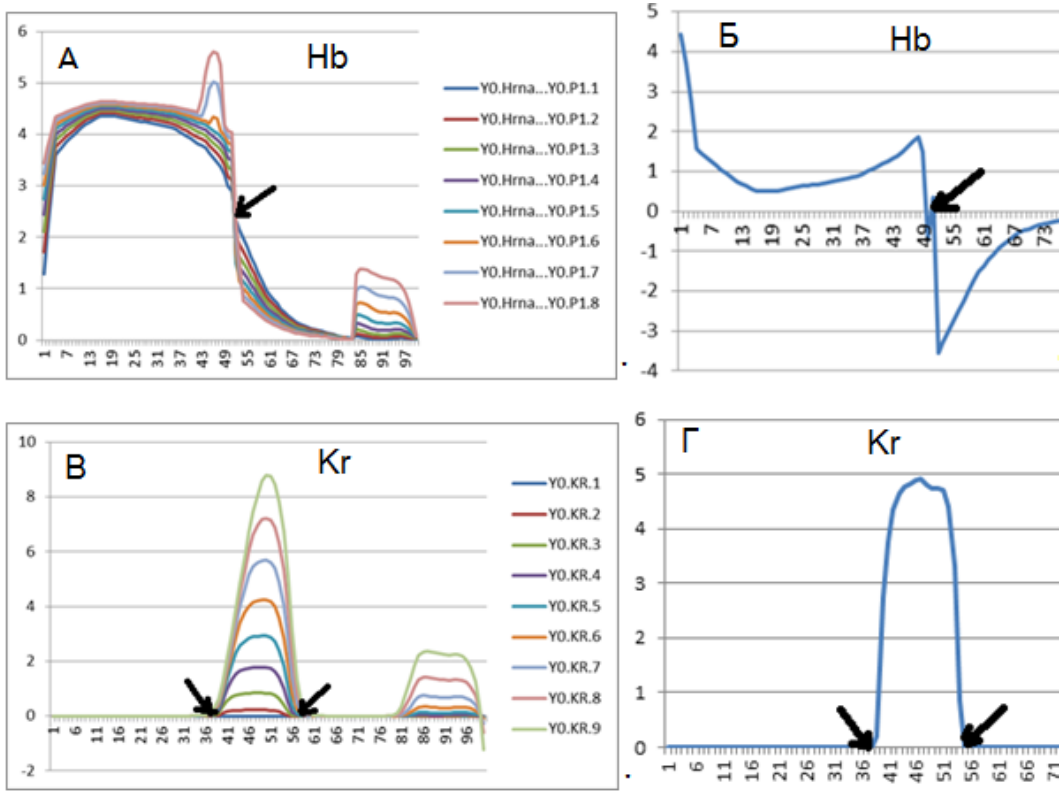

**Рисунок 3** — Динамика формирования паттернов экспрессии мРНК (А) Hb и (В) Kr. (Б) и (Г) Решения уравнения стационарности в момент t=0 для определения граничных точек (см. текст). Граничные точки указаны стрелками.

## 4. Позиционные характеристики модели

Рассмотрим теперь более внимательно механизмы формирования паттернов моделью и сосредоточимся на их характерных особенностях, имеющих биологическое значение.

Позиции границ доменов экспрессии. В процессе формирования паттернов генов gap позиции границ доменов экспрессии в передней части тела эмбриона постепенно устанавливаются, их крутизна растет, и сами домены становятся выше. Установление границ доменов имеет важное значение в развитии эмбриона, поскольку границы являются функциональными элементами, которые могут играть роль вторичных морфогенетических градиентов (входных факторов) для генов-мишеней, следующих за семейством gap в каскаде генов сегментации.

В работе [25] мы впервые обратили внимание на то, что при формировании пространственной экспрессии gap генов установление стабильных и крутых границ доменов связано с наличием в экспериментальном паттерне ядра, в котором в течение клеточных циклов 13-14A (NC13-14) экспрессия гена остается неизменной. Граница домена при этом формируется так, что к концу цикла крутой склон домена стабилизируется в позиции этого ядра, которое нами названо *граничным ядром* (или граничной точкой). Также нами показано, что в рамках коннекционистской модели [11] данная особенность формирования границ воспроизводится правильно, т.е. существует граничная точка, в которой уравнения модели стационарны:

$$R \cdot g[\Phi(TFs)] = \lambda P(t) \qquad \forall t, \qquad (7)$$

и положение граничной точки определяется в начальный момент времени.

Покажем, что в нашей модели также присутствуют граничные точки, хотя в силу нелинейности генных взаимодействий они могут иметь определенные отличия от коннекционистской модели.

Динамика формирования паттерна мРНК *hb* отвечает всем критериям формирования стабильной границы: присутствует начальный убывающий градиент, высота домена возрастает со временем и вне домена экспрессия равна нулю, при этом саморегуляция положительна и линейна $(\alpha_{Hb}^{(1)} > 0)$ [25]. Математически позиция граничной точки определяется из условия стационарности в начальный момент времени начальными условиями:

$$R \cdot g\left(\Phi(S_{Hb}^0, Gt, Kni, Nub)\right) = \lambda P_{Hb}^0, \tag{8}$$

где $P_{Hb}^0 = P_{Hb}^{(2)}(0)$ и $S_{Hb}^0 = S_{Hb}(0)$ — исходные паттерны мРНК и белка *hb*, соответственно, и учитывается, что $S_{Kr}^0 = S_{Kr}(0) \equiv 0$.

Стационарность означает, что $P_{Hb}(t) = const$ $\forall t$, т.е. л.ч. (7) также должна быть постоянна во времени. Для этого надо, чтобы выполнялось одно из двух следующих условий:

$$\begin{cases} \alpha_{Hb}^{(2)} S_{Hb}(t) + \alpha_{Bcd}^{(2)} F([Bcd]; K_{Bcd}) - \alpha_{Kr}^{(2)} F(S_{Kr}(t); K_{Kr}) - \alpha_{Kni}^{(2)}[Kni] - \alpha_{Gt}^{(2)}[Gt] \\ \quad - \alpha_{Nub}^{(2)}[Nub] \leq 0 \quad \forall t \\ P_{Hb}(0) = 0 \end{cases} \tag{9}$$

или

$$\alpha_{Hb}^{(2)} S_{Hb}(t) - \alpha_{Kr}^{(2)} F(S_{Kr}(t); K_{Kr}) = \text{const} \quad \forall t. \tag{9a}$$

При выполнении условия (9) в силу свойств сигмоиды левая часть (8) равна нулю, и тогда уравнение (1) стационарно в точке, где $P_{Hb}^{(2)}$ нулевое в течение всего периода моделирования. При выполнении (9a) в граничной точке имеет место динамическое равновесие концентраций белков $S_{Hb}(t)$ и $S_{Kr}(t)$ и стационарность концентраций мРНК $P_{Hb}(t)$. Именно последний сценарий реализуется в формировании границы hb, что показано на рисунке 3A.

В отличие от *hb*, мРНК *Kr* в начальный момент времени не экспрессируется, домен экспрессии начинает формироваться с нуля и его левая граница сразу же устанавливается, практически не смещаясь в процессе развития, в то время как правая несколько сдвигается прежде, чем занять окончательную позицию. При подобном формировании паттерна граничные точки существуют, но расположены в ядрах с нулевой экспрессией на границе формирующегося домена (см рисунок 3B.). Позиция этих точек определяется также условием, аналогичным (9):

$$\beta_{Kr} S_{Kr}(t) + \beta_{Hb} S_{Hb}(t) + \beta_{Hb2} S_{Hb}^2(t) + \beta_{Hb,Bcd} S_{Hb}(t) \cdot [Bcd] -$$
$$\beta_{Gt}[Gt] - \beta_{Kni}[Kni] - \beta_{Nub}[Nub] \leq 0 \quad \forall t. \tag{10}$$

В данном случае условие $P_{Kr}(0) \equiv 0$ по определению выполняется тождественно во всех точках паттерна, что вкупе с (10) сразу же влечет стационарность (7). При этом в момент $t = 0$ в (10) выполняется равенство и далее $\beta_{Hb} S_{Hb}(t) + \beta_{Hb2} S_{Hb}^2(t) + \beta_{Hb,Bcd} S_{Hb}(t) \cdot [Bcd]$ не возрастает во времени.

Таким образом, в качестве характеристических точек паттернов *hb* и *Kr*, мы рассматриваем граничные точки их доменов экспрессии, определяемые условиями (9-9a) и (10). Позиции этих точек приведены в первом столбце таблицы 1.

**Таблица 1.** Значение мер чувствительности граничных точек паттернов мРНК *hb* и *Kr*

| граничные точки | HB      | Kr      | Bcd      | Gt       | Kni      | Nub      |
|-----------------|---------|---------|----------|----------|----------|----------|
| Hb (52)         | 0.02439 | 0.00018 | 0.000636 | -0.00074 | -8.67E-06| -2.75281 |
| Kr (37)         | 0       | 0       | -0.35795 | -0.03367 | 0        | 0        |
| Kr (59)         | 0       | 0       | -0.31392 | 0        | -0.09784 | 0        |

## 5. Анализ чувствительности модели

Подгонка модели к данным обеспечивает, как было показано, хорошее соответствие моделируемых паттернов мРНК и белка экспериментальным. Важным вопросом остается, насколько результаты моделирования устойчивы к вариабельности входных данных. Важно исследовать, насколько чувствительны к пертурбациям входных факторов позиции характерных признаков паттернов, означенных в предыдущем разделе.

Позиция рассматриваемых нами характеристик определяется изначально пространственными паттернами входных факторов и далее не меняется. Такое поведение демонстрируют границы доменов *hb* и *Kr*, граничные точки на которых определяются уравнениями (7). Динамика формирования границ проиллюстрирована на рисунке 3. Чувствительность граничных точек в этом случае может быть исследована в начальный момент времени по начальным условиям модели.

Позиция граничных точек, в которых выполняется условие стационарности, определяется уравнением (8).

Чувствительность позиции к вариабельности как начальных условий $S^0$, так и неизменного по времени внешнего входного фактора *TF* определяется частной производной от л.ч. по концентрации данного фактора, вычисленная в стационарной граничной точке. Эта величина характеризует, насколько нарушится равенство в уравнении стационарности в случае изменения условий в начальный момент моделирования. Мера чувствительности к внешнему фактору *TF* тогда задается:

$$\begin{cases} 1/2 \cdot \frac{R}{\lambda}[1 + g(\Phi(TFs))] \cdot [1 - g(\Phi(TFs))] \cdot \partial\Phi/\partial[TF] & \text{при } \Phi(TFs) \geq 0 \\ 0 & \text{в против. сл.} \end{cases} \quad (11)$$

Здесь принимается во внимание известное свойство производной логистической функции: $g'(x) = 1/2 \cdot (1 - g(x))(1 + g(x))$. Вид функционала Ф, определяемый формулами (9-9a) и (10), дает ясное представление об относительной чувствительности позиции соответствующего признака к изменчивости концентраций разных входных факторов (в линейном случае просто определяемое коэффициентами при соответствующих концентрациях). Однако в силу нелинейности сигмоидной функции судить об абсолютной чувствительности можно только используя полную производную (11).

Чувствительность к изменениям начальных условий модели (а именно, концентрации моделируемых генов – *hb* и *Kr*) определяется

$$\begin{cases} 1/2 \cdot R/\lambda \, [1 + g(\Phi(TFs))] \cdot [1 - g(\Phi(TFs))] \cdot \partial\Phi/\partial[\text{Hb}] - 1 & \text{при } \Phi(TFs) \geq 0 \\ 0 & \text{в против. сл.} \end{cases}, \quad (12)$$

для производной по концентрации Hb, и аналогично для Kr.

Результаты приведены в таблице 1. Для границы *hb* можно сделать следующие выводы. Граница домена экспрессии мРНК чувствительна только к изменению дозы Nub и, в меньшей мере, самого исходного Hb. Остальные входные факторы особого влияния на локализацию граничной точки не оказывают. Границы Kr в начальный момент формируются под влиянием четырех факторов (Hb, Bcd, Gt и Kni), из которых только изменение дозы материнского Bcd может привести к сдвигу домена *Kr*.

# Заключение

Регуляция генов зачастую происходит сложно и требует моделирования с использованием подходов, выходящих за рамки простых линейных межгенных взаимодействий и основанных на принципах модулярности цис-регуляторных областей. В работе представлена модель пары генов *hb-Kr*, учитывающая реальную множественность ЦРМ у этих генов (биологически еще более реалистична). Модель получается методом компьютерного эволюционных вычислений (поэтому изначально, сразу подгоняется к экспериментальным данным). Модель включает кинетические члены второго и третьего порядка и в явном виде моделирующая кооперативность активации генов трансфактором бикоид и зависящее от концентрации действие трансфактора Kr на ген *hb* (при низкой концентрации — активатор, при высокой — репрессор).

.Выделены характерные особенности формирования паттернов экспрессии, имеющих биологическое значение. Нами определены математические принципы чувствительности позиции этих характеристик к возмущениям начальных условий и внешних факторов модели, постоянных во времени.

Метод позволяет выявить факторы, обеспечивающие позиционную устойчивость паттернов экспрессии МРНК и белка, что имеет важное значение в развитии организма.

## источники финансирования

# Список использованных источников